\documentclass[lettersize,journal]{IEEEtran}
\usepackage{amsmath,amsfonts}
\usepackage{algorithmic}
\usepackage{algorithm}
\usepackage{array}
\usepackage[caption=false,font=normalsize,labelfont=sf,textfont=sf]{subfig}
\usepackage{textcomp}
\usepackage{stfloats}
\usepackage{url}
\usepackage{verbatim}
\usepackage{graphicx}
\usepackage{cite}
\usepackage{xcolor}
\definecolor{rev_color}{RGB}{0,0,0}

\definecolor{rev2_color}{RGB}{15,170,0}

\begin{document}

\title{Opportunities and Challenges to Integrate \\Artificial Intelligence into Manufacturing Systems:\\ Thoughts from a Panel Discussion}

\author{Ilya Kovalenko, Kira Barton, James Moyne, and Dawn M. Tilbury 

\thanks{I. Kovalenko is with Department of Mechanical Engineering and Industrial \& Manufacturing Engineering, Pennsylvania State University, University Park, PA. {\tt\small \{iqk5135\}@psu.edu}.
K. Barton and D. M. Tilbury are with the Robotics Department and the Department of Mechanical Engineering, University of Michigan, Ann Arbor, MI, USA. {\tt\small \{tilbury, bartonkl\}@umich.edu}.
J. Moyne is with the Department of Mechanical Engineering, University of Michigan, Ann Arbor, MI, USA. {\tt\small \{moyne\}@umich.edu}.
}}

\maketitle

\section{Introduction}
\label{sec:intro}

\begin{figure}[t]
\centering
\includegraphics[width=1\columnwidth]{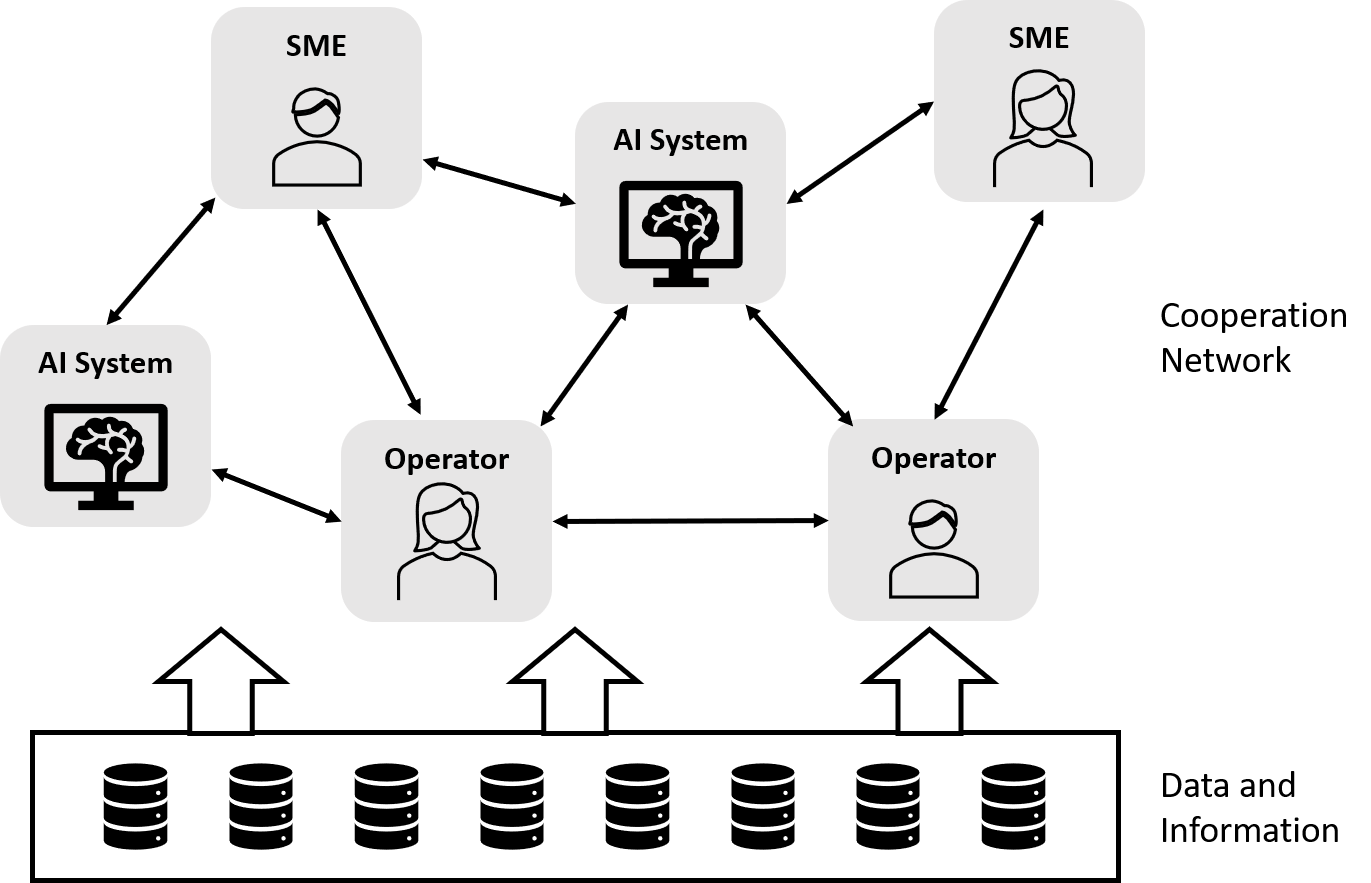}
\caption{Vision for integrating AI Systems into the manufacturing environment.}
\label{fig:vision}
\vspace{-0.5cm}
\end{figure}

Rapid advances in artificial intelligence (AI)\footnote{Caveat: The panel did not attempt to disentangle Artificial Intelligence from Machine Learning, and used the two terms loosely interchangeably during the discussion.} have the potential to significantly increase the productivity, quality, and profitability in future manufacturing systems. Traditional mass-production will give way to personalized production, with each item made to order, at the low cost and high quality consumers have come to expect. Manufacturing systems will have the intelligence to be resilient to multiple disruptions, from small-scale machine breakdowns, to large-scale natural disasters. Products will be made with higher precision and lower variability. While gains have been made towards the development of these factories of the future, many challenges remain to fully realize the vision shown in Figure~\ref{fig:vision}.

To consider the challenges and opportunities associated with this topic, a panel of experts from Industry, Academia, and Government was invited to participate in an active discussion at the 2022 Modeling, Estimation and Control Conference (MECC) held in Jersey City, New Jersey from October 3-5, 2022 \cite{mecc2022}. This panel discussion started with an overview presentation given by Professor Ilya Kovalenko (Pennsylvania State University) on the topic of automated learning for manufacturing systems \cite{kovalenko_toward_2022}. Following the overview, the panelists introduced themselves, their organizations, and provided preliminary thoughts on the integration of AI into manufacturing systems. Panelists included Dr. Meiling He (Rockwell Automation), Dr. Daewon Lee (Samsung AI Center NY), Dr. James Moyne (Applied Materials and University of Michigan), Dr. Robert Landers (University of Notre Dame), and Dr. Jordan Berg (National Science Foundation). The panel discussion focused on the challenges and opportunities to more fully integrate AI into manufacturing systems, and was moderated by Professors Kira Barton (University of Michigan) and Dawn Tilbury (University of Michigan).

Three overarching themes emerged from the panel discussion. First, to be successful, AI will need to work seamlessly, and in an integrated manner with humans (and vice versa). Second, significant gaps in the infrastructure needed to enable the full potential of AI into the manufacturing ecosystem, including sufficient data availability, storage, and analysis, must be addressed. And finally, improved coordination between universities, industry, and government agencies can facilitate greater opportunities to push the field forward. The rest of this article briefly summarizes these three themes, and concludes with a discussion of promising directions.

\section{Humans and AI working together}
\label{sec:human}

Although many large-scale manufacturing processes are automated to some degree, there are always humans involved at multiple levels. The humans may be physically interacting with the system (e.g., loading parts into a machine) or interacting through a computer (e.g., defining the production schedule). Different Subject Matter Experts (SMEs) have different expertise, some may have a deep technical knowledge of a specific process such as maintenance of a specific machine, while others may have a broader understanding of a system. Humans are expected to be adaptable, and able to react to both small disturbances (slightly out-of-order product flow) and large changes (new product arriving). 

Most AI deployed in manufacturing plants today is relatively fixed, i.e., it does not have the ability to learn beyond what it has been trained to do, nor can it ask questions when it does not understand what to do. The panel felt that there was a significant opportunity to develop AI that can work in a more integrated fashion with SMEs, at different levels, perhaps answering questions posed by the SME and even posing questions when something becomes uncertain. While some AI tools that take input from a human have been developed for data-science applications (e.g., Google VisionAI), these capabilities need to be extended to the manufacturing  environment. In this way, the knowledge held by the SMEs can be leveraged with the knowledge in the AI, and no human knowledge will be left behind. This integration of SME and AI can also lead to better trust of the AI by the SME, so that the AI is seen as a tool to help get the job done well, empowering the SME rather than threatening to eliminate the job.

Many currently-deployed AI systems in manufacturing detect anomalies that occur, things that are out of the ordinary, which can lead to preventing poor quality or machine breakdown. However, in many cases, the anomaly detection just alerts an operator, who is called over to investigate the situation, and make a decision about the appropriate response (ignore, shutdown, repair, etc). Most of these solutions are developed in a specific domain, using SMEs knowledgeable about the process. Generalizing these solutions to new domains is difficult, and generally requires new SMEs to be engaged. Could an AI be developed that would facilitate this type of generalization? What would it need to learn and where would it get the information?

In many manufacturing plants today, although robots and humans may work together to accomplish a job, they do not work in close proximity nor do they request knowledge from each other. They are separated due to safety constraints, with the robots sequestered in cages. Newer collaborative robots can operate safely around humans, although they are generally restricted to move slower than their caged counterparts. Despite closer proximity, collaborative robots do not request additional information from a human when new or unfamiliar challenges arise. As perception technologies improve, there are opportunities for AI to help robots and humans work together on physical tasks such as assembly. This could help address the challenge of personalized production.

As we move into the era of human-centric, sustainable manufacturing \cite{xu2021industry}, questions of cooperation between humans and AI become increasingly more important. One question that was raised was whether it was essential to differentiate between human and artificial intelligence. From a manufacturing system perspective, would it be possible to start looking at any intelligent system (humans, robots, computers, etc.) as an intelligent agent with specific capabilities? By blurring the line between human and artificial intelligence, we can start to develop an evolutionary process to support learning resulting from a continuum of human and AI knowledge.

\section{Infrastructure for AI in Manufacturing}
\label{sec:infrastructure}

Manufacturing systems are incredibly complex. From the nonlinear physics of low-level processes (welding, 3D printing) to the material handling within a plant, scheduling, and supply chain, everything needs to come together for a high-quality part to be produced and delivered to a consumer. There are hundreds of thousands of variables, of different types, and measured at different sample rates, that interact in diverse ways to describe what is happening. Many manufacturing plants operate multiple shifts per day, or even 24/7, leading to huge volumes of data. There are thousands or even millions of different things that can happen. AI technology needs data to learn, and although there is a lot of data to be found in manufacturing systems, it is not always AI-ready. Or, even though there is a lot of data, it may not be sufficiently diverse and capture enough samples of important events to develop effective learning and intelligence applications. In short, most data management systems were not developed or put in place with AI in mind (e.g., maintenance management databases), and so they are not AI ready in terms of many data quality dimensions (e.g., accuracy, freshness, granularity, volume/archive length, variety, and context richness).

Data is collected through sensors, and the panelists spent some time discussing the limitations of current sensing in manufacturing. In many cases, the sensors do not measure the variable of interest, limiting the ability for AI to learn a model of the process. For additive manufacturing or welding processes, the sensor can only measure what is on the surface and not the important material characteristics underneath. For some manipulation processes, only vision is available without any haptic or tactile feedback. Some sensors, such as thermal imaging cameras, can return huge amounts of data, which can be a challenge to process. 

AI problems require massive amounts of data, and the process of obtaining that data is expensive. Although manufacturing companies have a lot of data, they typically do not have the amount of data required for many AI approaches. Further, manufacturing companies are reluctant to share their data, due to its proprietary nature, reducing the potential for supplementing the data through additional resources. The panel discussed the advantages of industry-university partnerships to share information (e.g., data, pre-competitive model forms). Efforts to standardize the anonymization of information to preserve the underlying characteristics for different types of AI and learning would also be useful.

As discussed in Section~\ref{sec:human}, there is a big need to develop techniques that allow humans and AI technology to collaboratively work on tasks in the manufacturing system. However, the panelists also identified that there is a significant infrastructure challenge that needs to be addressed before humans and AI can effectively work together. Examples were provided where SMEs are expected to be both/either at the front end of the AI technology (e.g., setting up the bounds for anomaly detection) and/or the back end of the AI technology (e.g., analyzing the information provided by the AI tool). However, the interfaces may have been developed for a specific type of worker/SME. How can manufacturers start to develop generalizable interfaces that enable a comfortable working environment for the human? Can interfaces be developed that allow the human and AI to work on the tasks that each one is good at? How can AI be enhanced to the point of asking for information when necessary and accepting information asynchronously?

\section{Developing Industry-University Collaborations}

The panelists have experience in industry, academia, and government. The panel recognized that most of the identified challenges, from improving the AI infrastructure to developing new technology, cannot be solved by a single institution and must include collaborations between industry, academia, and the government. Companies will often have the technology and can identify real-world manufacturing problems that would benefit from the integration of AI. Universities provide the education infrastructure to develop the manufacturing workforce and the research capabilities that allow researchers to address various challenges. The government can foster these collaborations through programs that identify the need and provide funding for for AI in manufacturing. The need to incorporate the workforce (e.g., operators and SMEs) into the research discussions was also recognized. As AI technology is developed for the manufacturing sector, training tools for the workforce must be developed to improve the integration of new manufacturing AI technology. It will also be important to motivate the people working in the manufacturing environment to adopt and use AI technology for their work.

While AI has been recognized as a pivotal technology for growth and innovation for manufacturing \cite{deloitte_ai_2020}, industry has been slow to adopt AI technology within the manufacturing environment. Some of the factors that contribute to this slow adoption of AI technology include issues with the economy of scale. A lot of solutions to research problems are specific and cannot be scaled to the rest of the manufacturing enterprise due to cost and time constraints. As mentioned above, to advance AI solutions, there is a need for manufacturers to share data. Partnerships between universities and industry, with government support, can help with data and information sharing, such as developing improved data anonymization techniques and changing the cultural perspective of data sharing in manufacturing. 

\section{Promising Directions}
\label{sec:directions}

The panel discussed ideas about how to push the manufacturing field toward a more fully integrated paradigm of AI technology in manufacturing systems.  The advancement of manufacturing system technology (e.g, Industrial Internet of Things, collaborative robots) and the rise of customer expectations for customization and production capabilities provides an opportunity for the integration of AI in the manufacturing system industry. While a number of challenges were identified for the full integration of AI in manufacturing systems, there are a number of promising directions for this growing area. Industry, academia, and government agencies recognize the need for developing collaborations and programs that can facilitate improved adoption of AI in the manufacturing environment. Manufacturing scheduling and planning was one of the identified high-value, target research areas that can be targeted with AI technology. The panel believed that the scheduling and planning process can be improved through AI technology due to the significant amount of data available and the large opportunity for productivity improvement, especially when moving towards personalized production. The need for improving the reusability and extensibility of AI solutions via digital twin technology was also mentioned as a topic of interest \cite{moyne2020requirements}. Another key research area identified by all members of the panel was the need to integrate  subject-matter expertise into AI technologies deployed in the manufacturing domain at all levels. Both human and artificial intelligence are needed to realize the promise of high-quality, high-volume, low-cost personalized production systems. The discussion in this panel aligned with the research directions and opportunities that have been identified in existing literature. For example, the need to leverage technological advancements to help create human-centric and eco-centric manufacturing systems has recently been identified as a topic of interest by a number of other researchers in the area of smart manufacturing~\cite{leng_industry_2022,huang_digital_2021,xu2021industry,zhou_editorial_2022}. Note that while the panel discussion described in this manuscript covered a wide array of topics, there were a number of important research areas, such as the use of digitization and virtual manufacturing \cite{yang_parallel_2022} and the role of robots and cobots~\cite{evjemo_trends_2020}, that were only briefly mentioned and could benefit from future discussions on challenges and opportunities.

\section*{Acknowledgments}
We would like to thank the MECC organizers \cite{mecc2022} for the opportunity to present the special session.  We would also like to thank the panelists and audience for their thoughtful comments and participation in the session.

\bibliographystyle{IEEEtran}
\bibliography{magazine}

\vfill

\end{document}